\documentclass[12pt]{blois07}
\usepackage{graphicx,epsfig,amssymb}
\usepackage{cite,./mcite}

\begin{document}
\title{
{\small \rm DESY 07-197 \hfill SFB/CPP-07-78}\\
Status of Deeply Inelastic Parton Distributions}
\author{Johannes Bl\"umlein}
\institute{DESY, Platanenallee 6, D-15738 Zeuthen, Germany}
\maketitle

\begin{abstract}
A brief review on the status of unpolarized parton densities and the 
determination of the QCD scale $\Lambda_{\rm QCD}$ from deep-inelastic 
scattering data is presented. 
\end{abstract}

\section{Introduction}

\vspace{1mm} \noindent
Deeply inelastic lepton--nucleon scattering provides a clean way to 
extract the parton densities of the nucleons together with the QCD scale
$\Lambda_{\rm QCD}$. The exact determination of the parton densities is 
decisive 
for the understanding of the scattering cross sections at hadron colliders 
as LHC \cite{HLHC}. The main goal of the investigation is the measurement 
of the leading twist distributions. In the large $x$ region higher twist 
effects are measurable as well \cite{HT,BBG}.~\footnote{Similarly, one may 
hope to find higher twist effects in the region of small $x$ in the future
\cite{SATUR}.} During the last years the determination of moments of 
parton distribution functions \cite{LPDF} and the QCD scale 
\cite{ALPHa,ALPHb} 
within lattice-QCD calculations became more and more precise. A comparison 
of these results and the measurement of the corresponding quantities from 
precision data using higher order perturbation theory will provide highly 
non-trivial test of Quantum Chromodynamics. On the perturbative side, 
the running of $\alpha_s(Q^2)$ is known to 4--loop orders \cite{ALPH1} 
while the anomalous dimensions and the massless Wilson coefficients 
were calculated to 3--loop order \cite{THRL,WIL3}. The heavy flavor Wilson 
coefficients are known to 2--loop order only \cite{HEAV1,AB,HEAV11}. A 
first 
coefficient contributing at 3--loop order was calculated recently 
\cite{HEAV2}. Due to this the QCD analysis of deeply inelastic structure 
functions in $l^{\pm}N$ scattering may be performed for flavor 
non--singlet combinations to $O(\alpha_s^3)$ and to a very good 
approximation even to $O(\alpha_s^4)$, cf.~\cite{BBG}. In the flavor 
singlet case, strictly speaking, the analysis cannot be performed to 
3-loop order, since the corresponding heavy flavor Wilson coefficients 
are not known yet. It can be performed in an approximation to 3--loop 
order, describing the heavy flavor contributions to 2--loop order, which 
induces a remaining theoretical error. 
In the present paper we concentrate on the case of unpolarized 
deep-inelastic scattering. A recent overview on the status of polarized 
parton densities was given in \cite{POL1}. The paper is organized as 
follows. In section~2 we summarize main aspects of QCD analyzes and 
discuss recent progress in measuring unpolarized parton distribution 
functions. Section~3 summarizes determinations of $\Lambda_{\rm QCD}$ 
in deeply inelastic scattering and in Section 4 we discuss future 
perspectives.
\section{QCD Analysis of Unpolarized Structure Functions}

\vspace{1mm} \noindent
In case of light--cone dominance the deeply inelastic 
structure functions at twist--2 are described by a Mellin convolution of
the bare parton densities and the hard scattering cross sections, which 
are both infinite, but are renormalized to finite parton densities and
Wilson coefficients by absorbing the ultraviolet singularities of the 
latter into the former~:
\begin{eqnarray}
F_j(x,Q^2) &=& \hat{f}_i(x,\mu^2) \otimes
\sigma^i_j\left(\alpha_s, \frac{Q^2}{\mu^2}, x \right)\\
& &  \hspace{0.5cm} \uparrow {\rm {\tiny bare~pdf}}~~~~~~\uparrow {\rm
{\tiny sub-system ~cross-sect.}}
\nonumber\\
\hspace{2cm}
&=&
\underbrace{{\hat{f}_i(x,\mu^2)} \otimes
{\Gamma^i_k\left(\alpha_s(R^2), \frac{M^2}{\mu^2},
\frac{M^2}{R^2}\right)}}_{{\large \rm finite~pdf} \equiv f_k}
\otimes
\underbrace{{C^k_j\left(\alpha_s(R^2), \frac{Q^2}{\mu^2},
\frac{M^2}{R^2},x
\right)}}_{{\large\rm finite~Wilson~coefficient}}
\nonumber
\end{eqnarray}
\normalsize
The scale evolution of the structure functions
is described by the 
Symanzik-Callan equations for the ultraviolet singularities\cite{REN},
and likewise for the renormalized parton densities and Wilson 
coefficients, 
\begin{eqnarray}
\left[
M \frac{\partial}{\partial M} + \beta(g)
\frac{\partial}{\partial g} - 2 \gamma_{\psi}(g)
\right] F_i(N) &=& 0
\\
\left[M \frac{\partial}{\partial M} + \beta(g)
\frac{\partial}{\partial g} + \gamma_\kappa^N(g) - 2 \gamma_{\psi}(g)
\right] f_k(N) &=& 0 \\
\left[M \frac{\partial}{\partial M} + \beta(g)
\frac{\partial}{\partial g} - \gamma_\kappa^N(g)
\right] C_j^k(N) &=& 0~. 
\end{eqnarray}
Here the Wilson coefficients contain as well the heavy quark degrees of 
freedom, while the parton distributions can only be defined for strictly 
massless partons in the respective kinematic region, i.e. for collinear 
particles. Clearly for $Q^2 \lessapprox m_H^2$ heavy quarks cannot be 
treated as partons. It is known for long \cite{EHLQ} that the heavy quark 
contributions have quite different scaling violations if compared to light 
partons for a very large range in $Q^2$.
 
The solution of the evolution equations is easiest 
being performed in Mellin space. Here the corresponding evolution 
equations can be solved to all orders in the coupling constant 
analytically, cf. e.g.~\cite{BV}. The solution has to be continued 
analytically from even values of the Mellin moment $N \rightarrow 
N~\epsilon~{\bf C}$. This requires the continuation of harmonic 
sums \cite{HSUM} representing the higher order anomalous dimensions and 
light flavor 
Wilson coefficients \cite{ANCONT} 
and that of heavy flavor Wilson coefficients \cite{AB}. 
At every loop order and expansion depth in the dimensional regularization 
parameter $\varepsilon$ a uniform maximal number of basis elements is 
needed to construct the respective single--scale quantities.
To 3--loop orders 14 basic Mellin transforms are sufficient \cite{STR1}.
The structure of 
this representation is characterized by meromorphic functions in the 
complex $N$ plane, the perturbative part of it obeys nested recursions $z 
\rightarrow z-1$ and can be constructed analytically starting from the 
respective asymptotic representation in the region $|z| \rightarrow 
\infty$, cf.~\cite{STRUCT}. The expression for the structure functions 
used in the $\chi^2$-minimization can be easily obtained by a single fast 
numeric contour integral around the singularities of the problem. To keep 
the evolution code fast all relations expressing the evolution kernels can 
be stored in large arrays during the initialization of the code, while in 
the minimization procedure only the parameters of the parton distribution 
functions are varied along with $\Lambda_{\rm QCD}$. The procedure can be 
systematically generalized including resummations, e.g. in the small--$x$ 
region \cite{BV}. These effects, however, were found to be non-dominant in 
the region of HERA data. Initially large effects are likely canceled by 
sub-leading terms almost completely, as being the case for all quantities 
calculated in fixed 
orders up to $O(\alpha_s^3)$. Usually three sub-leading terms (series) are 
required to obtain the correct result, cf. \cite{BV}. We will therefore 
not include effects 
of this kind in the present analysis, see \cite{JBREV} for a survey.
Other recent analyzes also find only small small $x$ effects \cite{SX} in 
the 
evolution of $F_2(x,Q^2)$ in the region $x \gtrapprox 10^{-4}$ currently 
probed at HERA.

\hspace*{-1cm}
\begin{minipage}[t]{\linewidth}
\centering\epsfig{figure=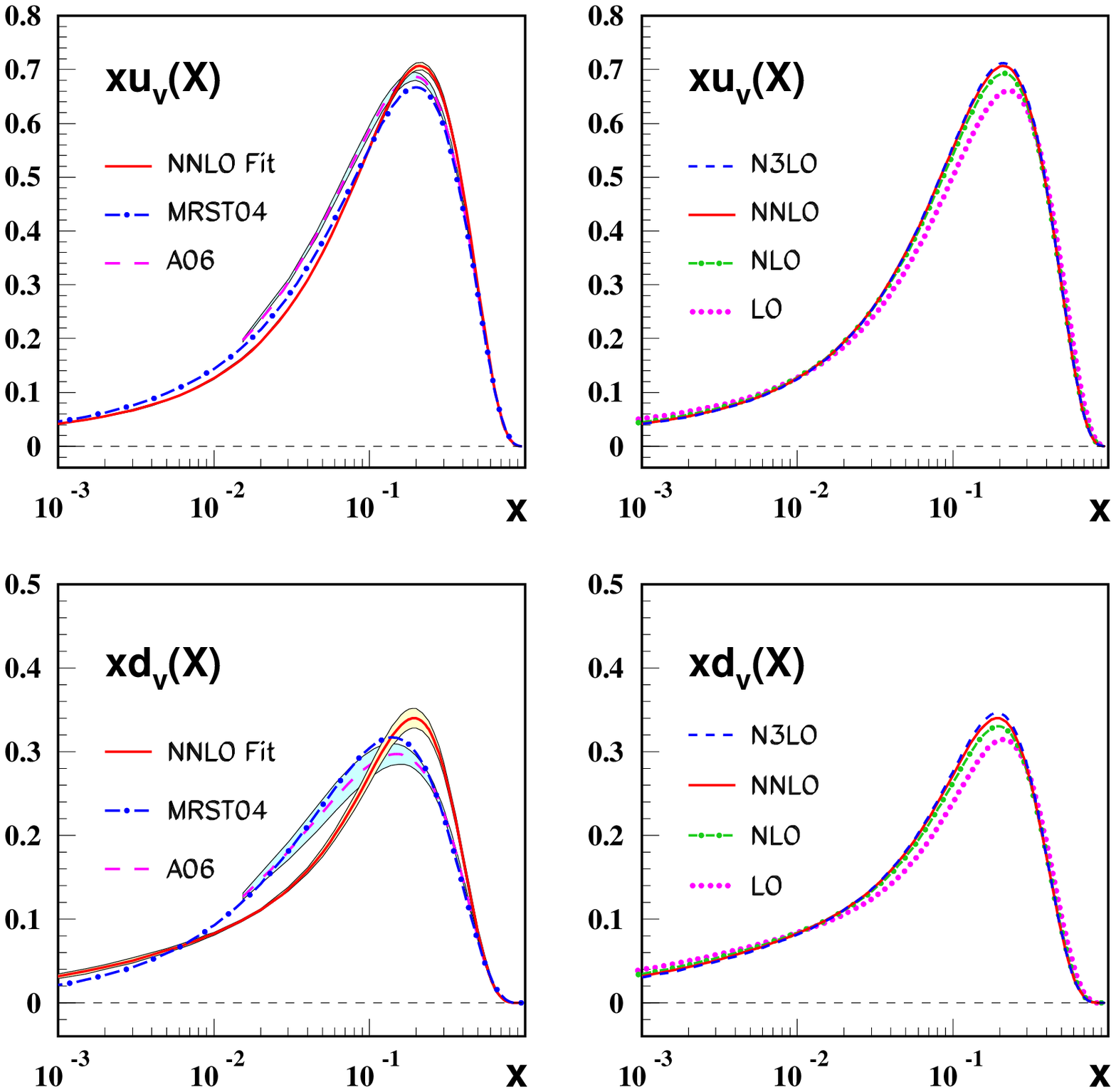,width=0.6\linewidth}\\
{\small Figure~1: The NNLO valence quark distributions, \cite{BBG}, 
compared to other analyzes and perturbative stability of the fit comparing
different higher order corrections.}
\end{minipage}

\vspace{2mm}
A flavor non--singlet analysis of the deep-inelastic world data was 
carried out recently in \cite{BBG}. This analysis primarily aimed on 
measuring $\alpha_s(M_Z^2)$ widely free of gluonic effects. Due to the 
fact that the $O(\alpha_s^3)$ Wilson coefficients dominate the scaling 
violations at the 4--loop level and the effect of the splitting function 
is rather minor only,  as estimated by a Pad\'{e}-approximation, the 
analysis 
is 
effectively of 4--loop order. We accounted for a $\pm$ 100\% error in the 
estimated 4-loop anomalous dimension. Comparison with the second moment of 
the non-singlet 4--loop anomalous dimension \cite{BC} showed agreement
within better than 20 \% well confirming our error treatment.
In Figure~1 the fit results are shown for the valence quarks and compared 
to other analyzes \cite{MRST,AMP} (left figures). The right figures show
the convergence of the analysis from leading order (LO) to 4-loop order
(NNNLO).

In Ref.~\cite{BBG} also a model-independent extraction of higher 
twist-contributions in the large $x$ region was performed. Here it is 
essential to describe the leading twist contributions as accurately as 
possible, since the leading twist Wilson coefficients are large in the 
large $x$ region. 

The light sea quark densities are known at lower precision if compared to 
that of the valence quarks. Here still more data are required. The 
distribution $x(\overline{u} - \overline{d})(x,Q^2)$ can be obtained 
from Drell--Yan data. In a recent analysis \cite{AMP} improved sea quark 
distributions $x(\overline{u} \pm \overline{d})$
were obtained, see Figure~2a. The 3--loop corrections 
lower the theory error to the level of the experimental accuracy. A recent 
determination of the strange quark density was performed by the CTEQ 
collaboration \cite{CTEQs}, see Figure~2b. This distribution is about half 
the value of that of the up and down sea quarks. 

\hspace*{-1cm}
\begin{minipage}[t]{\linewidth}
\centering\epsfig{figure=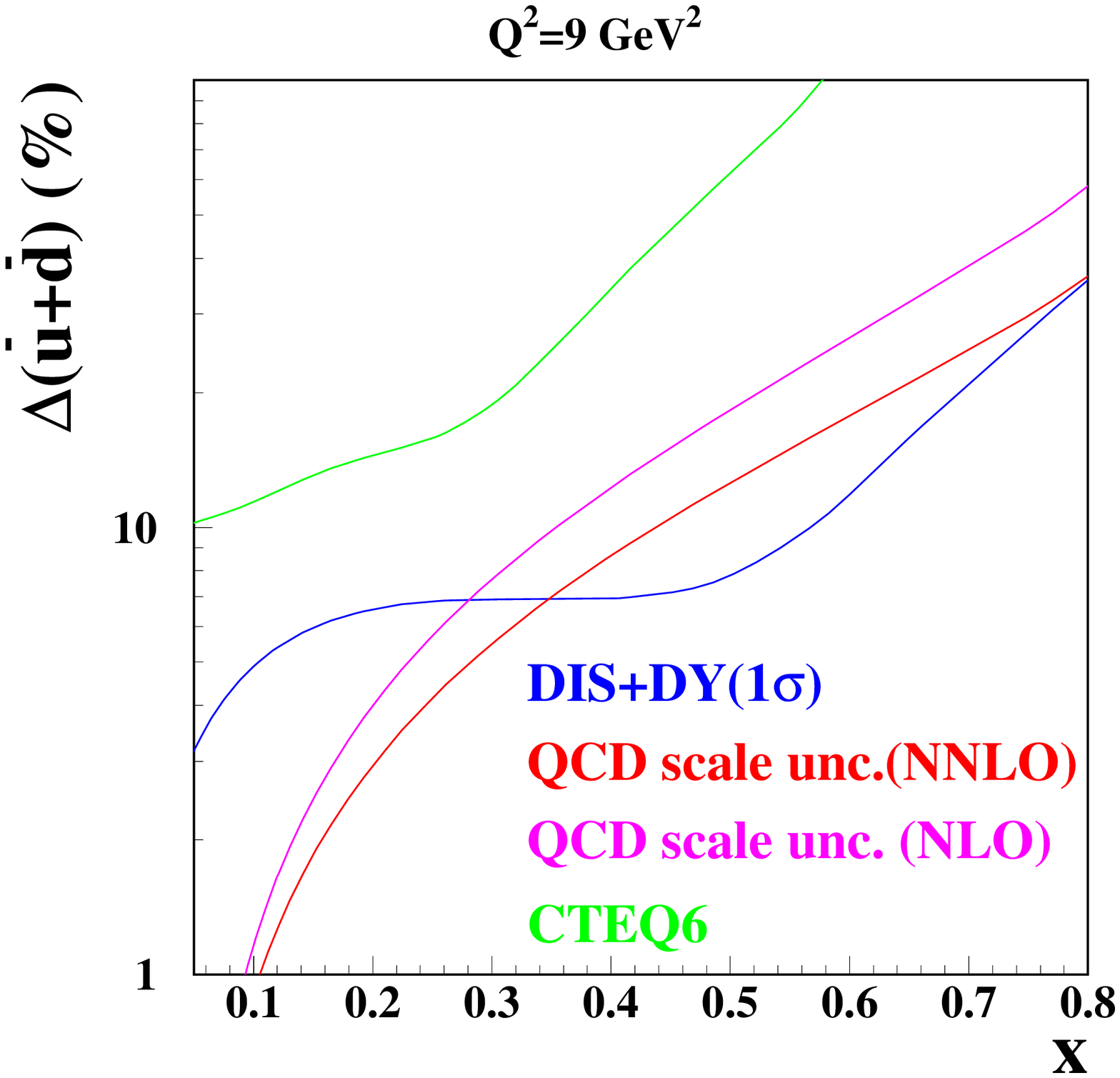,width=0.48\linewidth}
\centering\epsfig{figure=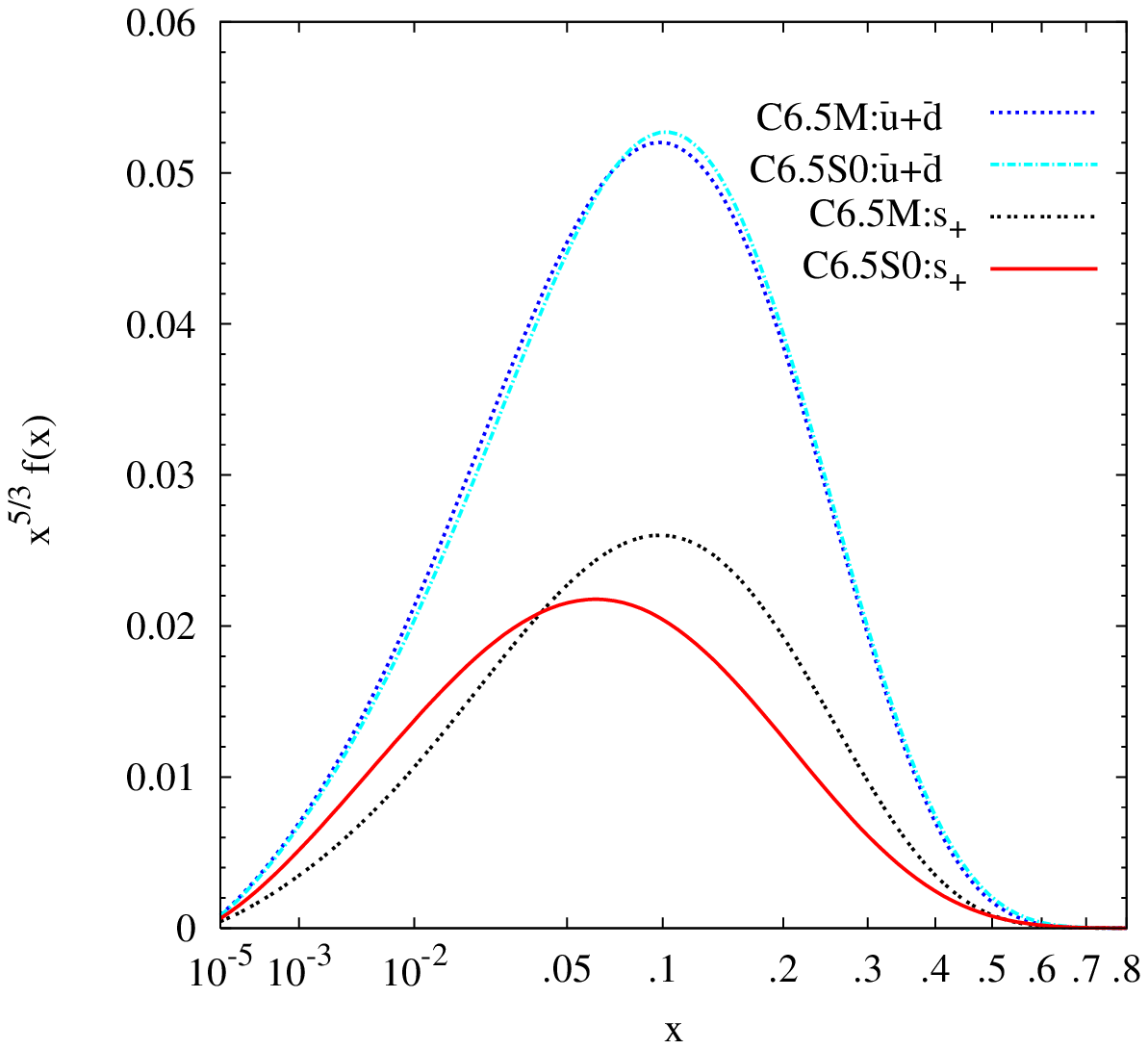,width=0.48\linewidth}

\vspace{1mm}\noindent
{\small Figure~2:~Uncertainty of $x(\overline{u}+\overline{d})$ 
distribution 
\cite{AMP} (left). The light flavor distributions for $Q^2 = 1.69 
{\rm GeV}^2$, 
Ref.~\cite{CTEQs}.}
\end{minipage}

\vspace{2mm}
The correct determination of the gluon density is of central importance 
since many scattering processes at LHC are gluon induced. The gluon 
distribution is rapidly growing as $x \rightarrow 0$ with rising values of 
$Q^2$. This expectation is confirmed by different analyzes 
\cite{AMP,GR,CTEQ1}. As an example we show the results of the recent 
analysis \cite{GR} in Figure~3a, where a rising behaviour is found down to 
scales of $Q^2 = 2~{\rm GeV}^2$. In contrast to this MSTW  \cite{MRST1} 
find a 
gluon distribution which is turning to lower values in the region $x 
\approx 10^{-3}$ for scales $Q^2 = 5~{\rm GeV}^2$ and lower, contrary to 
the 
results found in \cite{AMP,GR,CTEQ1}. 
The value of $\alpha_s(M_Z^2)$ in \cite{MRST1} $0.1191 \pm 0.002 \pm 
0.003$  
comes out larger than 
that in \cite{BBG,AMP,GR}, $\alpha_s(M_Z^2) = 0.1142 \pm 0.0021; 
0.1128 \pm 0.0015; 0.112.$ In \cite{CTEQ1} a determination of $\alpha_s$ 
is not undertaken, since the different data sets used in the fit bear too 
different systematics to allow this, which was outlined in \cite{CTEQ2} in 
detail. The analysis in \cite{MRST} differs from that in \cite{AMP} due to 
the inclusion of jet data from Tevatron, which are known to require a 
larger value of $\alpha_s(M_Z^2)$. 

\hspace*{-1cm}
\begin{minipage}[t]{\linewidth}
\centering\epsfig{figure=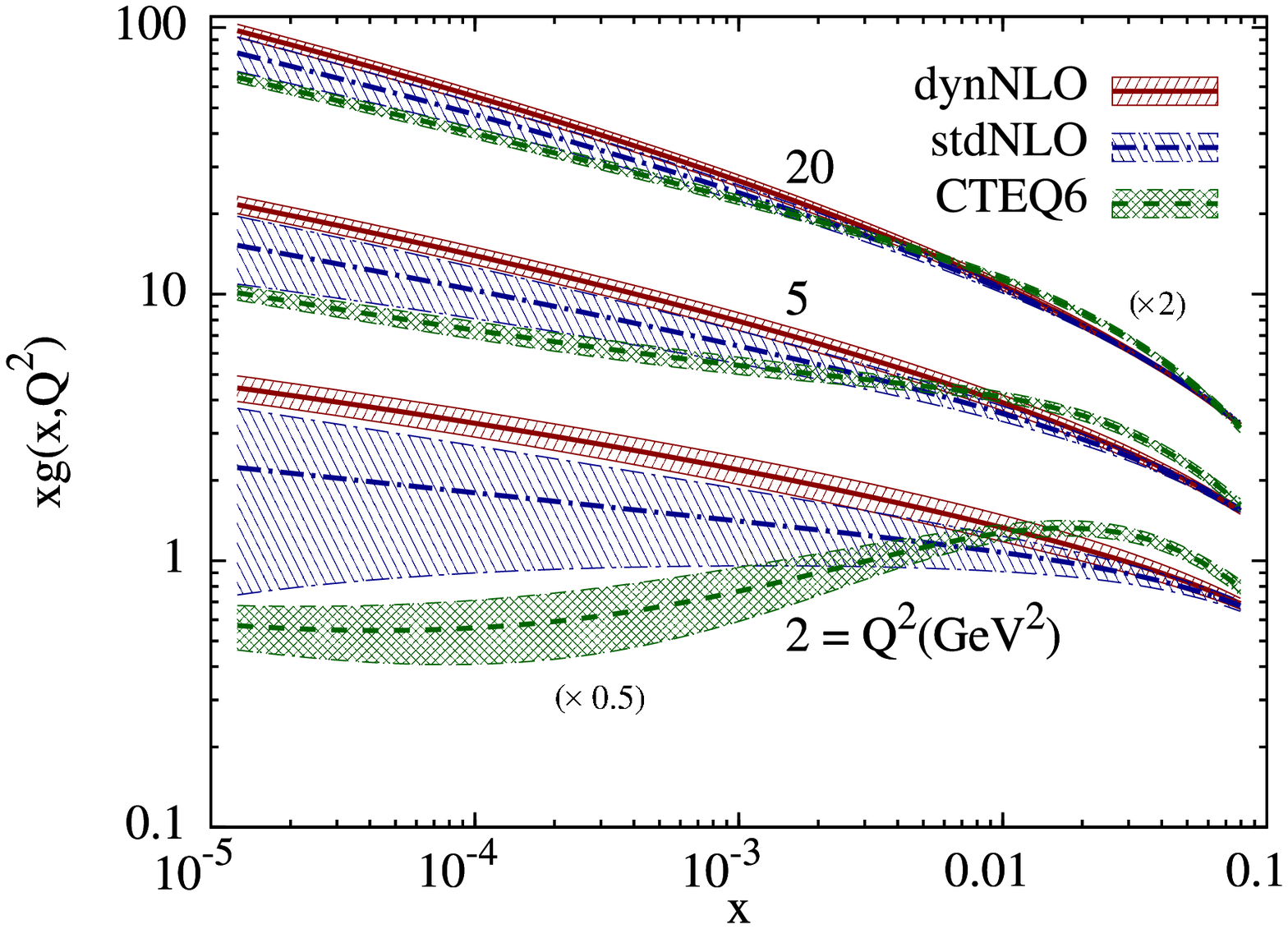,width=0.60\linewidth}
\centering\epsfig{figure=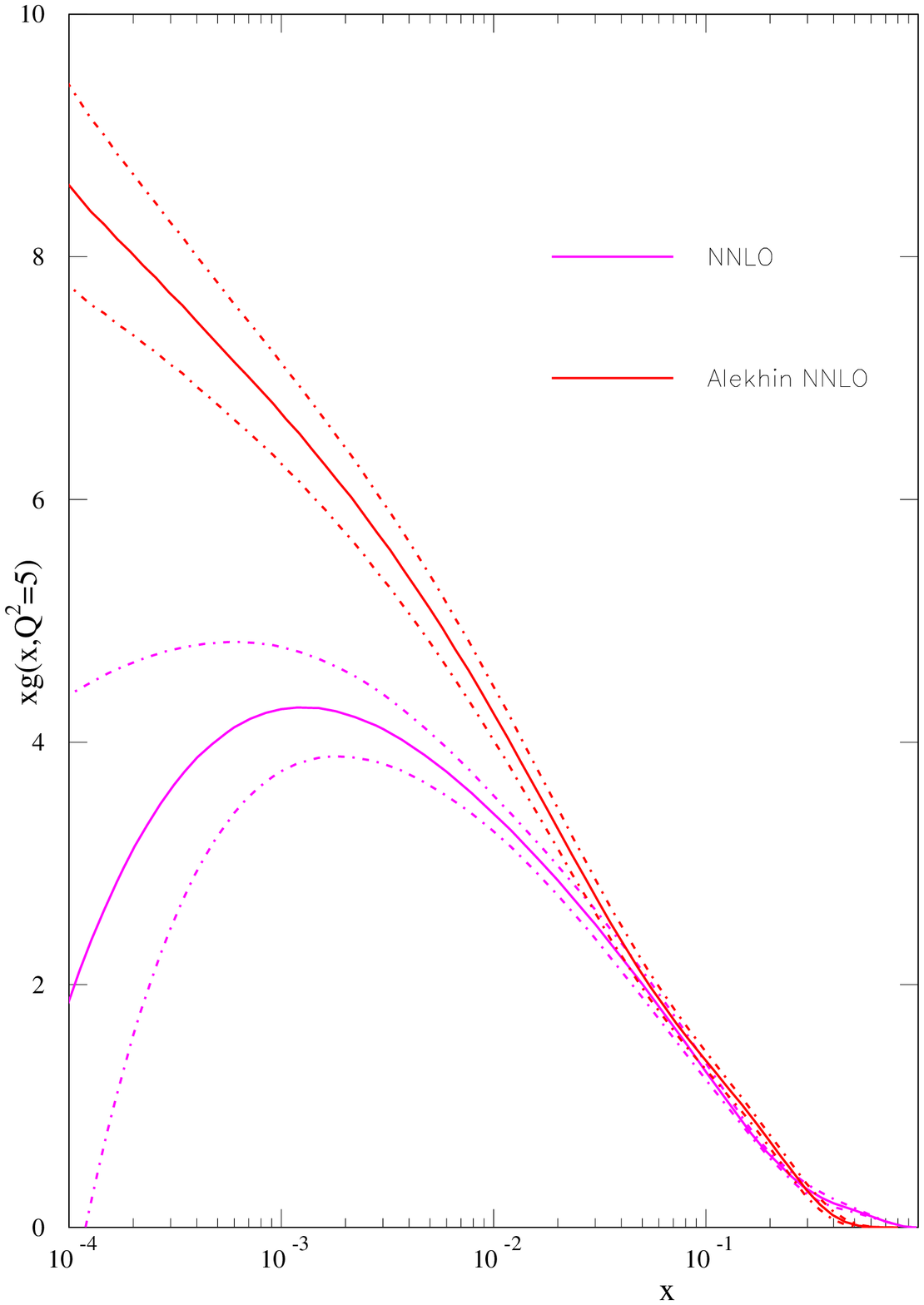,width=0.35\linewidth}

\vspace{1mm}\noindent
{\small Figure~3~: Gluon momentum distribution at NLO \cite{GR,CTEQ1} 
(left) and 
at NNLO \cite{MRST,AMP} (right).}
\end{minipage}

The measurement of $F_L(x,Q^2)$ can help to clarify this question. A 
recent analysis \cite{GR1} shows very good agreement with the current 
measurements \cite{FL}, which are partly still preliminary. The question 
of the correct value of the gluon distribution function should be 
clarified soon.  
\section{\boldmath $\Lambda_{\rm QCD}$ and $\alpha_s(M_Z^2)$}

\vspace{1mm} \noindent
A summary on different measurements of $\alpha_s(M_Z^2)$ from $l^{\pm} N$ 
scattering data in NLO, NNLO, and NNNLO is given in Figure~4, see also 
\cite{Blumlein:2007dk}. Present analyzes are carried out at the 3-loop 
level based on the anomalous dimensions \cite{THRL} and Wilson 
coefficients \cite{ZN}. If the analysis is restricted to deeply inelastic 
data the values of $\alpha_s(M_Z^2)$ come out somewhat lower as the world 
average \cite{SIGI}.
The convergence of the perturbative extraction of 
$\alpha_s(M_Z^2)$ out of the deeply--inelastic world data 
\cite{BBG} is illustrated comparing the central values from NLO 
to NNNLO~:
\begin{equation}
\alpha_s(M_Z^2) = 0.1148 \rightarrow 0.1134 \rightarrow 0.1142 \pm 0.0021.
\end{equation}
The change from the N$^2$LO to the N$^3$LO value is found deeply inside
the current experimental error. The  N$^3$LO value corresponds to
\begin{equation}
\Lambda_{\rm QCD}^{\rm \overline{MS},
N_f =4} = 234~\pm 26~{\rm MeV}.
\end{equation}

\hspace*{-1cm}
\begin{minipage}[t]{\linewidth}
\centering\epsfig{figure=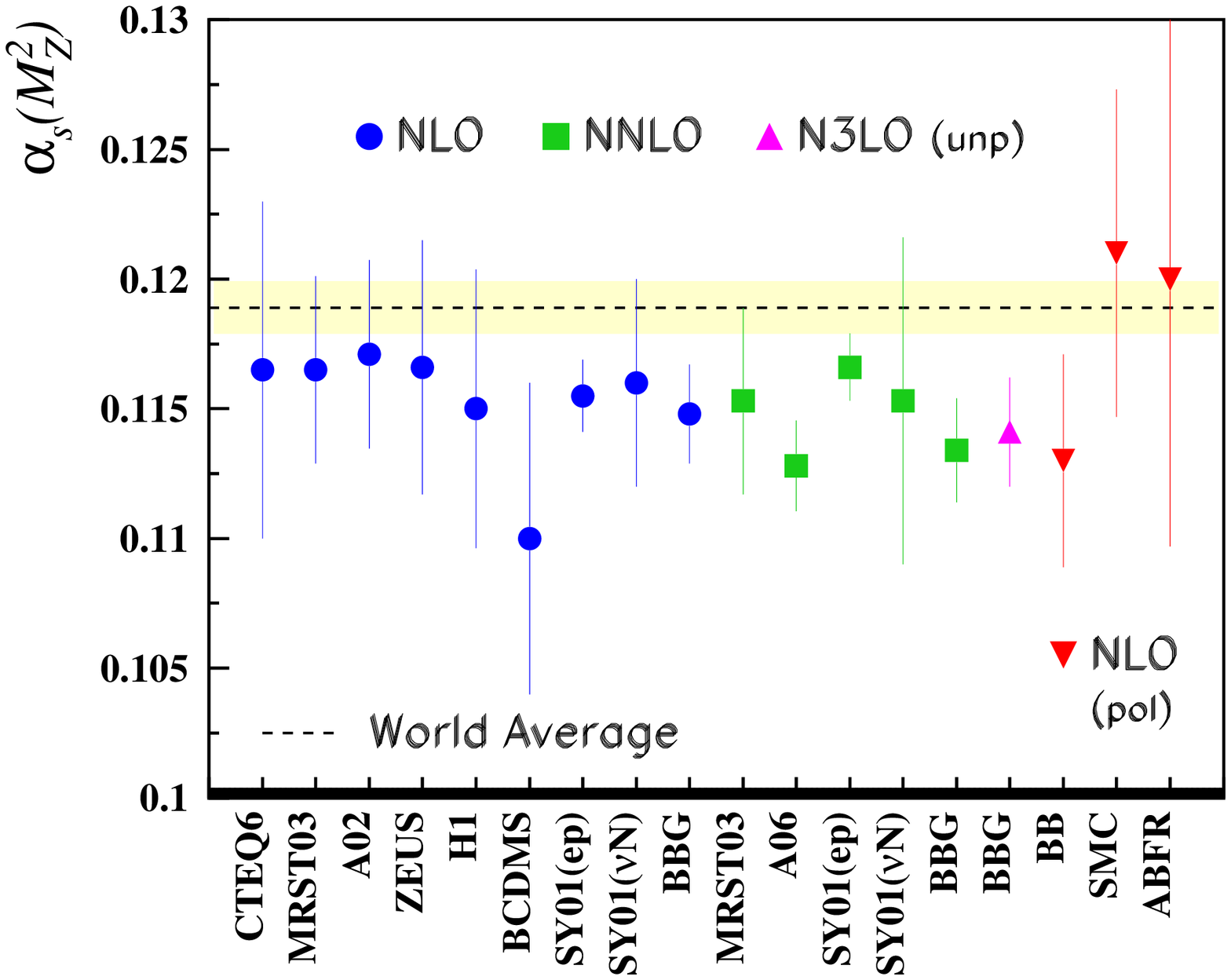,width=0.66\linewidth}

\vspace{1mm}\noindent
{\small Figure~4:~Summary of $\alpha_s(M_Z^2)$ measurements in 
deep-inelastic 
unpolarized and polarized $l^{\pm} N$ scattering: from NLO to NNNLO,
Ref.~\cite{BBG}.}
\end{minipage}

$\Lambda_{\rm QCD}^{\overline{\rm MS}}$ was measured also in two recent
lattice simulations based on two active flavors ($N_f =2$). These
investigations paid
special attention to non-perturbative renormalization and kept the
systematic errors as small as possible.
\begin{eqnarray}
\Lambda_{N_f=2}^{\rm latt} = 245 \pm 16 \pm 16~{\rm 
MeV}~~[6],~~~~~~
\Lambda_{N_f=2}^{\rm latt} = 261 \pm 17 \pm 26~{\rm MeV}~~[7]~.
\end{eqnarray}
A direct comparison with the case $N_f = 4$ in the
above data analyzes is not yet possible. However, the
difference between the earlier $N_f = 0$ and the present result in
$\Lambda_{\rm QCD}$ amounts to $O(10~{\rm MeV})$
only. We have to wait and see what is obtained for $N_f = 4$ in coming
analyzes.
{\section{Future Perspectives}}

\vspace{1mm} \noindent
Most of the data taken at HERA still have to be analyzed to extract the 
final data of {$F_{2,L}(x,Q^2)$}, {$F_2^{Q\overline{Q}}(x,Q^2)$}, 
and other structure functions. 
The analysis of these  measurements will be mandatory 
for the final 
precision determination of the parton distribution functions in the small 
$x$ region, in particular for the gluon and sea quark distribution 
functions. 
Important 
informations on the large $x$ behaviour of the valence quark densities will 
be obtained from JLAB \cite{ENUP}. Currently our knowledge of the 
individual light flavor sea quark distributions is still rather 
limited.
Here, the measurement of the Drell-Yan process and $W^{\pm}$ and 
$Z$--production at LHC will add in significant further information. That 
far signs of non-linear gluon evolution were not found in deeply inelastic 
scattering, unlike suggested by earlier theoretical expectations 
\cite{SATUR}. As the scale at which these effects come into operation 
cannot be determined pertubatively one has to search for those effects at 
still smaller values of $x$ using suitable scattering cross sections 
at LHC in the near future. After the completion of the HERA programme 
still inclusive measurements at much higher luminosity are required 
to determine some of the parton densities at higher precision. For a 
more detailed measurement of the sea quark distributions deuteron targets 
are required at high luminosity~\cite{JB87}. Here a programme like 
foreseen for the EIC \cite{EIC} can contribute essentially. The flavor 
contents of the sea-distribution can be analysed in great detail at 
high luminosity neutrino factories operating at higher 
energies \cite{NUFACT}. The results of both these facilities will be 
instrumental to explore distributions, which are more difficult to 
access as the polarized distribution functions, the transversity 
distribution, as well as the twist--3 and higher twist correlation 
functions to perform further rather non-trivial tests of QCD also in this area.
Various of these observables can be accessed at high precision in lattice 
calculations in the near future. In this way ab-initio predictions at 
the one side can be compared to  precision data analyzed within perturbation 
theory to higher orders on the other side. It is therefore  highly 
desirable, that these facilities \cite{EIC,NUFACT} are built in the 
future. 

\vspace{1mm}
\noindent
{\bf Acknowledgement.}\\
This work was supported in part by DFG Sonderforschungsbereich
Transregio 9, Computergest\"utzte Theoretische Physik.

\begin{footnotesize}

\end{footnotesize}
\end{document}